# Resolution enhancement of placenta histological images using deep learning

**Arash Rabbani[1,2], Masoud Babaei[3]**
[1]School of Mathematics & Statistics, University of Glasgow, Glasgow G12 8QQ, United Kingdom
[2] School of Computing, University of Leeds, Leeds LS2 9BW, United Kingdom
[3]Department of Chemical Engineering, University of Manchester, Manchester M1 3AL, United Kingdom
a.rabbani@leeds.ac.uk; masoud.babaei@manchester.ac.uk

***Abstract -*** In this study, a method has been developed to improve the resolution of histological human placenta images. For this purpose, a paired series of high- and low-resolution images have been collected to train a deep neural network model that can predict image residuals required to improve the resolution of the input images. A modified version of the U-net neural network model has been tailored to find the relationship between the low resolution and residual images. After training for 900 epochs on an augmented dataset of 1000 images, the relative mean squared error of 0.003 is achieved for the prediction of 320 test images. The proposed method has not only improved the contrast of the low-resolution images at the edges of cells but added critical details and textures that mimic high-resolution images of placenta villous space.

## 1. Introduction
Placenta is a vital organ that develops during the pregnancy to provide a medium for exchanging oxygen and nutrients between mother and fetus [1]. Micro-structure of internal placental tissues is an important feature in developing a healthy fetus [2]. These tissues have been widely investigated by sampling chorionic villus via biopsy and preparing histological slides [3]. However, sometimes histological images do not have either the required quality suitable for pathological examinations or cell counting. This study presents a simplified approach to improve the quality of the placenta histological images using deep learning. Although deep learning has been widely used for analysis [4] and quantification of placenta histological images, quality improvement techniques for these images have not been yet practiced.

## 2. Material and Method
### 2.1. Image dataset
In this study, deep artificial neural networks have been used to improve the quality of the histological placenta images. Required data for training and test purposes have been acquired from the virtual microscopy database [5] donated by the University of Michigan. The utilized H&E-stained images [6] belong to two healthy volunteers and are sub-sampled from the fetal zones where villous space is visible. Images have been captured at 40X magnification in which many different cell types are clearly visible and distinguishable. In these images, boundaries of blood cells are clearly visible as well as villous membrane cells. However, this is not the case in many of the publicly available dataset of human placenta images. To train a quality improvement model, we deliberately decrease the resolution of our high-resolution dataset and use it as a feed for a deep learning structure. For this purpose, cubic interpolation has been implemented to resample the images into a quarter of their original size. Then, using the same approach, images are brought back to the original size, but with a considerable amount of loss in details in the transformation process. Such artificial low-resolution images are assumed to mimic original low-resolution images that have been captured using a lower magnification or when adequate visual definition is lacking. A dataset of 32 images, (16 from each of the healthy volunteers) with the size of 1024×1524 have been augmented by random cropping and flipping to 1320 images with the size of 512×512 pixels. From this dataset, 1000 images have been used for training and 320 images have been used for test purposes.

## 2.2. Model structure

The deep learning model employed in this study is a tailored version of the U-net [7] with 4 down-sampling and 4 up-sampling steps. At each step, two consecutive convolutions with the filter size of 3×3 and same-size padding are used with *Exponential Linear Unit* [8] activation function and *He Normal* kernel initializer [9]. Depth of the convoluted layers increases from 16 to 256 as multiplied by 2 in each of the down-sampling steps. In addition, a 2×2 max-pooling filter with the stride size of 2×2 is used to shrink the size of the image in each of the mentioned steps.

Input and output images have the same dimensions of 512×512 pixels with 3 colour channels of red, green, and blue. The output image is designed to be the pixel-wise difference between the low- and high-resolution images. It is noteworthy that for practical convenience, the pixel values of the output residual image are shifted to be positive integers between 0 and 255, with the value of 127 representing no change and it is identified by grey colour. Accordingly, darker shades than the dominant grey denote a reduction in the average intensity of pixels and lighter shades represent a required increment of the pixel intensities. Fig. 1 illustrates the structure of the deep learning model, input and output images, and a sample reconstruction of a high-resolution image by pixel-wise addition of the low-resolution image and the predicted residual.

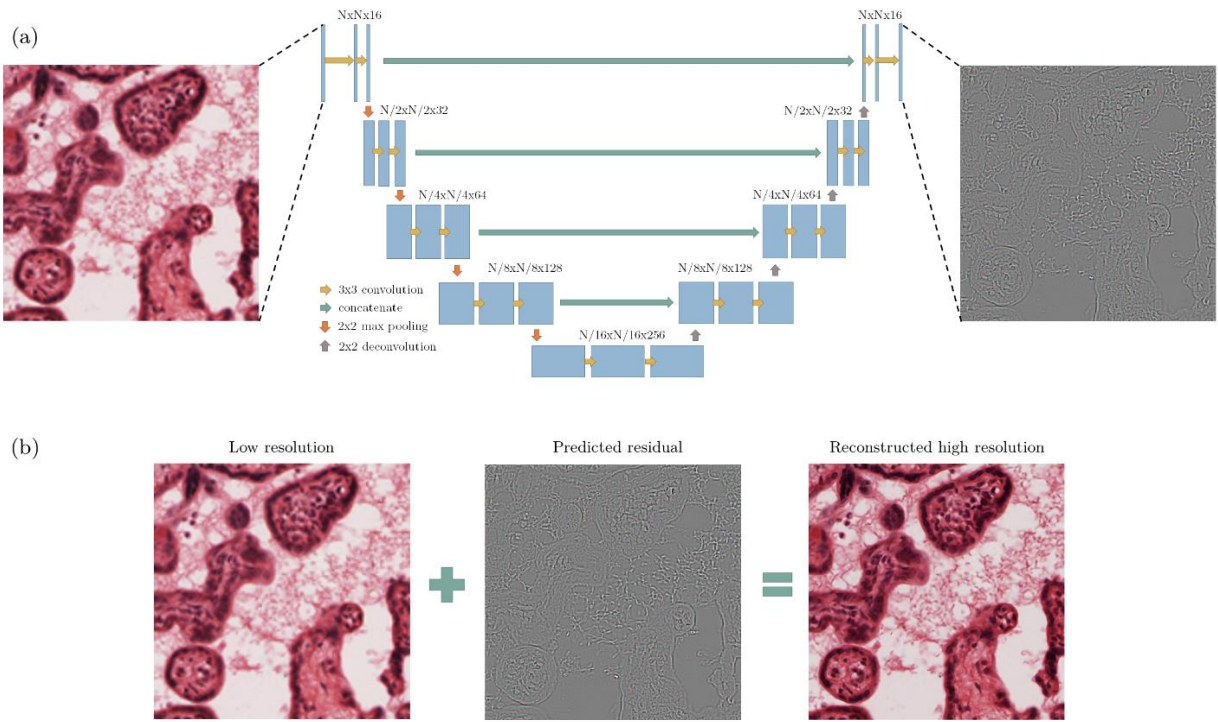

Fig. 1: Structure of the deep learning model including sample input and output images (a), an example of reconstructing a high-resolution image using a predicted residual image by pixel-wise summation (b). (N is the dimension of the input image in pixels and equal to 512.)

## 3. Results and conclusions

Using the proposed method, the described deep learning model is trained for 900 epochs with a batch size of two. The stopping criterion for this training is "no improvement in training accuracy after 100 epochs". The loss function used for training is binary cross-entropy [10] with a learning rate of 0.001, minimized via the *Adam* optimization technique [11]. The relative mean square error when predicting the training and testing datasets are 0.002 and 0.003, respectively. The trained model and related developed codes are written in Python using the Tensorflow package with Keras back-end and are available at a GitHub repository. (www.github.com/ArashRabbani/PlacentaSR).

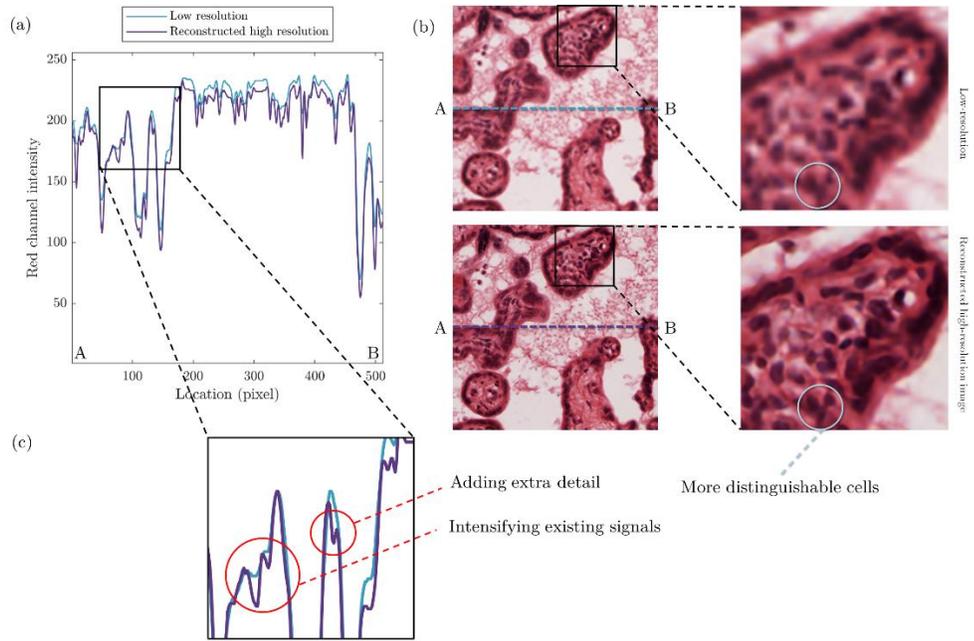

Fig. 2: An example of resolution improvement with plotted red channel intensities from point A to B on both low-resolution and reconstructed high-resolution images.

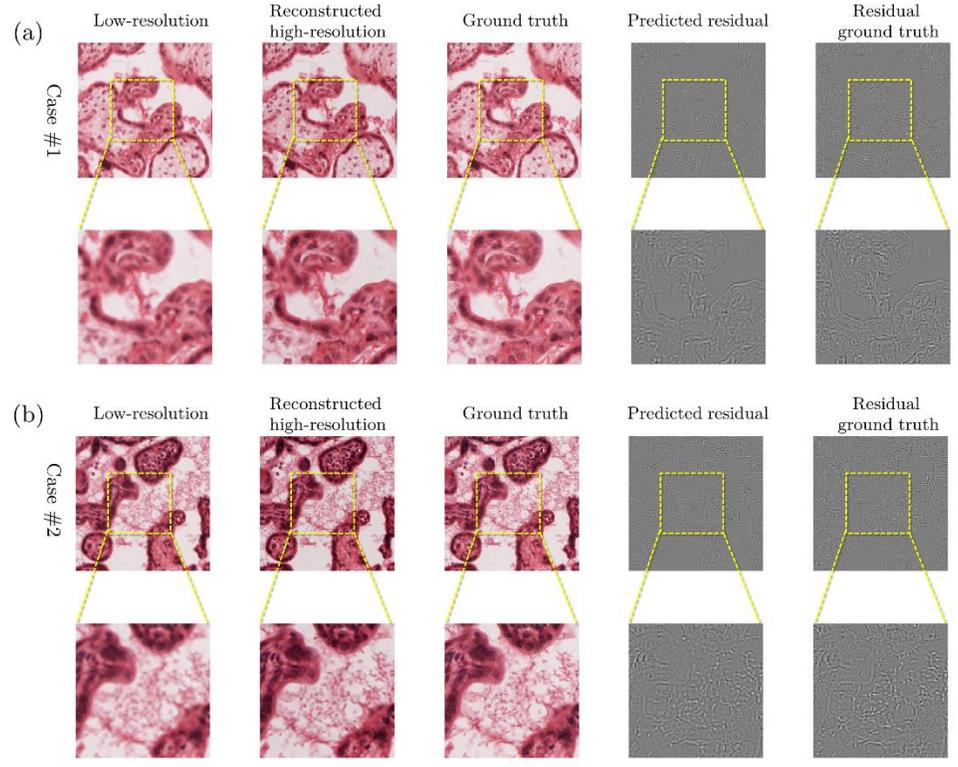

Fig. 3: Two examples of prediction from test dataset including, low-resolution image, reconstructed high-resolution image, grand truth high-resolution image, predicted residual and residual ground truth. The central section of all images has been magnified for better visualization of the slight differences.

Fig. 2 represents a sample reconstruction of a high-resolution image based on a low-resolution input in addition to visualization of the red channel intensity variations. As it can be seen, when moving from point A to B on the graph, the red channel intensity of both images follows the same pattern, but with two differences. The reconstructed image has more intensified signals compared to the low-resolution image. In addition, some extra details have been added to the high-resolution image by the deep learning model that gives a better representation of the texture of the placenta villous space. In the reconstructed high-resolution image, cells are more distinguishable and consequently more suitable for pathological analysis if needed.

As another visual presentation, Fig. 3 illustrates two examples of image resolution improvement based on two test images from two different healthy volunteers named cases #1 and #2. As it can be seen in both cases, image quality has improved to the level that the reconstructed high-resolution image is almost identical to the ground truth image based on a visual comparison. Also, the predicted residual is adequately similar to its ground truth.

Based on the achieved resemblance between the images, the presented method in this study is suggested to be used prior to manual or machine-based analysis of placenta histological images, especially in the cases where image quality is a limitation. In our studied cases, the trained model can improve the image quality approximately from 20X magnification to 40X magnification. The method does not only sharpen the edges and intensified the weak signals but also creates local details and textures within the villous space as they exist in an original high-resolution image.

## Acknowledgment

The authors thank the University of Manchester for the President's Doctoral Scholarship Award 2018 granted to Arash Rabbani to carry out part of this research.